\documentclass{aa}
\usepackage{graphicx,fixltx2e}
\usepackage{txfonts,hyperref}
\usepackage{natbib} 
\bibpunct{(}{)}{;}{a}{}{,}
\usepackage{color}
\definecolor{myRed}{rgb}{0.84,0.08,0.52}
\definecolor{black}{rgb}{0,0,0}
\definecolor{blue}{rgb}{0,0,1}
\usepackage[scaled=0.92]{helvet}
\usepackage[modulo,switch]{lineno}

\usepackage{hyperref}
\hypersetup{
   final=true,
   pageanchor=true,
   colorlinks=true,
   breaklinks=true,
   linkcolor=blue,
   citecolor=blue,
   urlcolor=blue,
   pdfpagemode=UseNone,
   pdftitle={Evidence of quiet Sun chromospheric activity related to an emerging small-scale magnetic loop},
   pdfauthor={Gömöry, P., Balthasar, H., Puschmann, K.G.},
   pdfsubject={Astronomy and Astrophysics},
   pdfkeywords={Sun: magnetic topology – Sun: surface magnetism – Sun: photosphere – Sun: chromosphere}}

\begin{document}

   \title{Evidence of quiet Sun chromospheric activity related to an emerging 
          small-scale magnetic loop}

   \author{P. G\"{o}m\"{o}ry\inst{1},
           H. Balthasar\inst{2},
           \and
           K. G. Puschmann\inst{2}
          }
   \authorrunning{P. G\"{o}m\"{o}ry et al.}

   \institute{Astronomical Institute of the Slovak Academy of Sciences,
              SK–05960 Tatransk\'{a} Lomnica, Slovakia\\
              \email{gomory@astro.sk}
         \and
             Leibniz-Institut f\"{u}r Astrophysik Potsdam, An der 
             Sternwarte 16, D–14482 Potsdam, Germany\\
             \email{hbalthasar@aip.de, kgp@aip.de}             
             }

   \date{Received March 05, 2013; accepted June 14, 2013}

 
  \abstract
   {}
   {We investigate the temporal evolution of magnetic flux emergence in the quiet 
   Sun atmosphere close to disk center.}
   {We combine high-resolution SoHO/MDI magnetograms with TRACE observations 
   taken in the 1216\,\AA~channel in order to analyze the temporal evolution 
   of an emerging small-scale magnetic loop and its traces in the chromosphere.}
   {At first place, we find signatures of flux emergence very close to the edge 
   of a supergranular network boundary located at disk center. The new emerging  
   flux appears first in the MDI magnetograms in form of an asymmetric bipolar 
   element, i.e. the patch with negative polarity is roughly two-times weaker 
   than the corresponding patch with opposite polarity. The average values of 
   magnetic flux and magnetic flux densities reach 1.6$\times 10^{18}$\,Mx, 
   -8.5$\times 10^{17}$\,Mx, and 55\,Mx\,cm$^{-2}$, -30\,Mx\,cm$^{-2}$, 
   respectively. The spatial distance between the opposite polarity patches 
   of the emerged feature increases from about 2\farcs5 to 5\farcs0 during 
   the lifetime of the loop which was not longer than 36\,min. A more precise 
   lifetime-estimate of the feature was not possible because of a gap 
   in the temporal sequence of the MDI magnetograms. The chromospheric response 
   to the emerged magnetic dipole occurs $\sim$9\,minutes later with respect to 
   the photospheric magnetograms. It consists of a quasi-periodic sequence of 
   time-localized brightenings visible in the 1216\,\AA~TRACE channel apparent 
   for $\sim$14 minutes and being co-spatial with the axis connecting the two 
   patches of opposite magnetic polarity.}
   {We identify the observed event as a small-scale magnetic loop emerging at photospheric 
   layers and subsequently rising up to the chromosphere. We discuss the possibility 
   that the fluctuations detected in the chromospheric emission probably reflect 
   magnetic field oscillations which propagate to the chromosphere in form of waves.}


   \maketitle
%

\section{Introduction}

Magnetic structures emerging in form of small-scale loops in the quiet Sun 
atmosphere and associated dynamics have recently come to the center 
of attention. Systematic studies of these features became possible mainly 
through space-born observations acquired with a new generation of instruments 
like the spectro-polarimeter 
\citep[SP;][]{2001ASPC..236...33L}               
of the Solar Optical Telescope 
\citep[SOT;][]{2008SoPh..249..167T}              
on board the Japanese space mission \textit{Hinode}
\citep{2007SoPh..243....3K}                       
or the Imaging Magnetograph eXperiment 
\citep[IMaX;][]{2011SoPh..268...57M}              
of the balloon-borne observatory \textit{SUNRISE} 
\citep{2010ApJ...723L.127S}                       
as they provide extended time-series of data with very high spatial resolution which is 
needed for the loop detection 
\citep[see e.g. works of][]{2007ApJ...666L.137C,  
                            2009ApJ...700.1391M,  
                            2010ApJ...713.1310I,  
                            2010ApJ...723L.185W,  
                            2010ApJ...714L..94M,  
                            2011ApJ...730L..37M,  
                            2012ApJ...755..175M,  
                            2012ApJ...745..160G,  
                            2012ApJ...746..182O,  
                            2012A&A...537A..21P,  
                            2012ApJ...747L..36V}. 
However, 
\citet{2007A&A...469L..39M}                        
and 
\citet{2010A&A...511A..14G}                        
showed that also observations obtained with ground-based instruments which are equipped 
with an adaptive optics system [e.g., the German Vacuum Tower Telescope 
\citep[VTT;][]{1985VA.....28..519S}               
with the Kiepenheuer Adaptive Optic System 
\citep[KAOS;][]{2003SPIE.4853..187V}]             
are suitable to study processes related to small-scale loop emergence. 

Small-scale loops represent a significant fraction of the magnetic flux in the quiet 
photosphere
\citep{2007A&A...469L..39M}                       
and are thus important for a more complex description of the latter.
\citet{2009ApJ...700.1391M}                        
performed an extensive statistical analysis of 69 emerging loops and found that 16 of 
these features even reach the transition between the upper photosphere and the lower 
chromosphere and that 10 of them can be seen yet in emission in \ion{Ca}{ii}\,H 
filtergrams. But, limitations in the used datasets did not allow them to identify any 
loop in the layers sampled by H$\alpha$ emission. Such evidence was given by 
\citet{2010ApJ...722.1970Y}                        
who observed the trailing part of an active region and found that even a very small 
dipole (spatial extent of $\sim$0\farcs5) can rise high enough and heat up sufficiently 
to be detected in the upper chromosphere where it exhibits significant activity. 
%
   \begin{figure*}
   \centering
   \includegraphics[width=16.5cm]{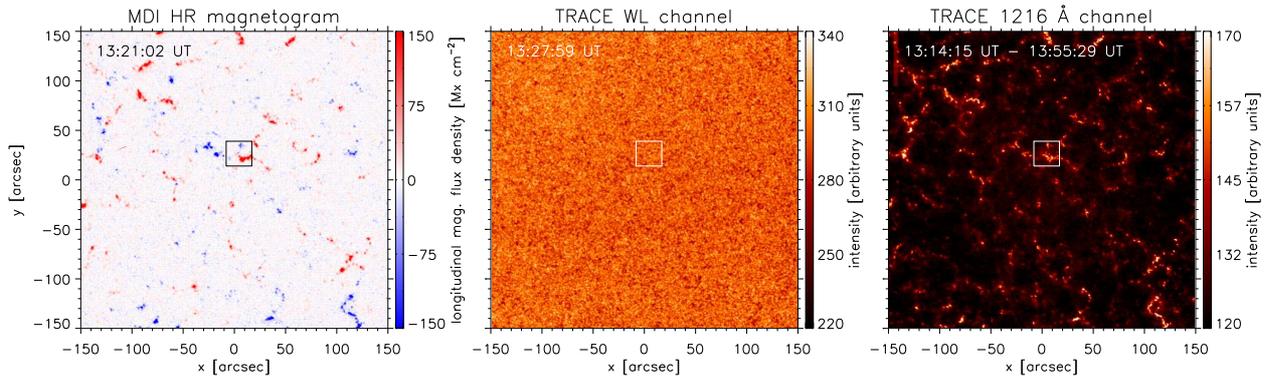}
   \caption{Context images showing the common FoV for all observations. 
            {\em Left panel}: SoHO/MDI magnetogram at best visibility 
            of the emerging feature. The color scale is in Mx\,cm$^{-2}$ 
            with red (blue) indicating positive (negative) polarity. An 
            artificial saturation at $\pm$150\,Mx\,cm$^{-2}$ has been 
            chosen to highlight regions of weak magnetic polarity on the 
            solar surface. 
            {\em Middle panel}: Image taken in the TRACE white-light channel 
            during the emergence of the analyzed feature demonstrating the 
            absence of strong magnetic fields in the FoV (quiet Sun). 
            {\em Right panel}: Temporal average of TRACE images taken in the 
            1216\,\AA~channel during the feature emergence showing long-living 
            chromospheric network elements. Squares enclose the region enlarged 
            in Fig.\,\ref{loop_evol}. (A color version of this figure is 
            available in the online journal.)}
            \label{context_image}
    \end{figure*}
%

These observational results support findings of 
\citet{2004Natur.430..326T}                        
who argued that the magnetic energy stored in the quiet photosphere is sufficient to 
balance the radiative losses of the chromosphere. Similar results were found by 
\citet{2008A&A...481L..25I}                        
for plage regions. Moreover, MHD simulations of 
\citet{2008ApJ...679L..57I}                        
suggest that small-scale loops that emerge in the photosphere can really reach 
chromospheric heights and get reconnected with the local expanding vertical magnetic 
fields, thus heating the surrounding chromospheric plasma. All these facts assign 
an important role to small-scale magnetic loops in transport and dissipation of magnetic 
energy for heating the chromosphere. However, different mechanisms capable for the 
transfer of energy into upper photospheric layers have still to be affirmed, especially 
with respect to results showing a lack of total energy flux transported to the chromosphere 
by acoustic waves 
\citep{2007PASJ...59S.663C},                      
although newer results obtained by
\citet{2009A&A...508..941B,2010ApJ...723L.134B}    
point out that the acoustic energy flux might have been underestimated in the past due 
to insufficient spatial resolution. In contrast, 
\citet{2013A&A...553A..73B}                        
suggest that magnetic heating processes are more important for the chromospheric energy 
balance than commonly assumed, when they compare photospheric magnetic fields obtained 
from spectro-polarimteric observations in the \ion{Fe}{i}\,630.25\,nm line of the 
Polarimetric Littrow Spectrograph 
\citep[POLIS;][]{2005A&A...437.1159B}             
with results retrieved after the application of a recently developed LTE-inversion strategy 
\citep{2013A&A...549A..24B}                       
on POLIS \ion{Ca}{ii}\,H spectra.  

However, while some simulations show that emerging loops can reach the chromosphere, 
simulations of magneto-convection by 
\citet{2006ApJ...642.1246S}                        
predict that these loops should disintegrate as they rise through the lower solar 
atmosphere, thus an energy transport connected to these features should be implausible. 
Moreover, a large variety of chromospheric dynamics can be modeled with a wave-driven 
reconnection 
\citep[e.g.][]{2009ApJ...701L...1D}.              
The latter could indicate that the small-scale emerging fields are at least not 
the only contributor to the chromospheric energy balance.

The discrepancies mentioned above do not allow to make any general conclusions in this 
research field so far, thus emphasizing the need of further observational studies.
A complete understanding of magnetic flux emergence in quiet Sun could considerably 
improve not only our knowledge about the photosphere but also shed light on the problem 
of chromospheric heating. 

In this work, we present a case study of an emerging small-scale loop that appeared 
close to a supergranular network boundary at the disk center and exhibited significant 
chromospheric activity. 

\section{Data and data reduction}

The analyzed data-set was obtained within the Joint Observing Program JOP 171 
which was performed during several days in the second half of October 2005. 
This observing program was dedicated to study properties of the quiet solar 
atmosphere from photospheric layers up to the corona.\footnote{JOP 171 details: http://sohowww.nascom.nasa.gov/soc/JOPs/jop171} To do so, several instruments on board the Solar and 
Heliospheric Observatory \textit{(SoHO)} and the Transition Region And Coronal 
Explorer \textit{(TRACE)} were involved in the data acquisition process. 
%
   \begin{figure*}
   \centering
   \includegraphics[width=16.5cm]{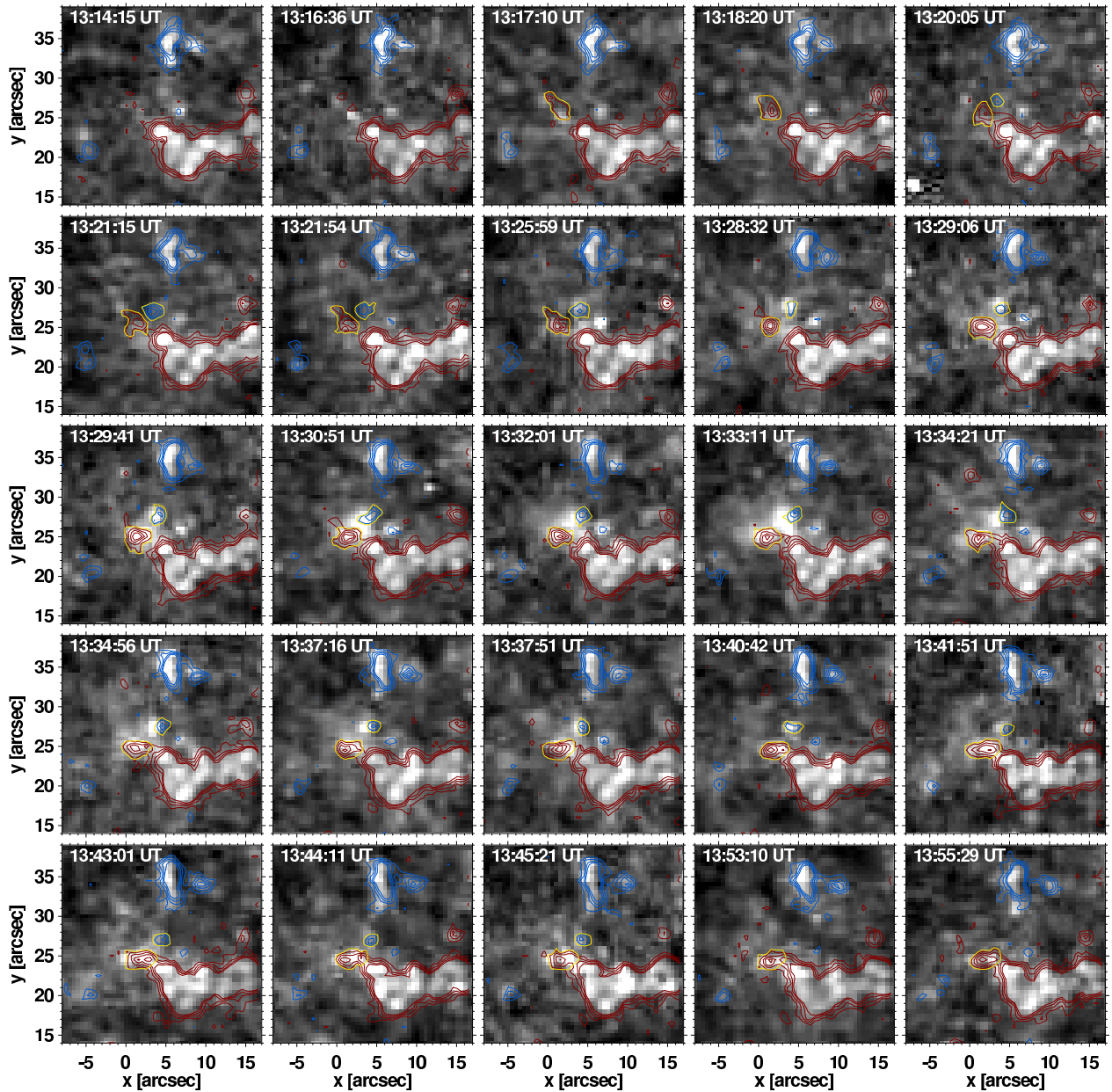}
   \caption{Temporal evolution of the emerging magnetic dipole. Background 
            images have been taken in the \textit{TRACE} 1216\,\AA~channel. 
            The superimposed contours represent the magnetic signal detected 
            in the \textit{SoHO}/MDI data. The magnetic field contours are 
            plotted at $\pm$25,  $\pm$40,  $\pm$55 and $\pm$70\,Mx\,cm$^{-2}$ 
            with red (blue) contours referring to positive (negative) magnetic 
            polarity. The yellow contours mark the pixels which were 
            used for the calculation of both the magnetic flux density and the 
            total magnetic flux at the footpoints of the emerging loop under 
            study. The recording time of the particular 1216\,\AA~images 
            is presented at the upper right corner of each sub-panel. The 
            corresponding magnetograms are almost co-temporal (with a maximal 
            time difference $<$\,30\,s). (A color version of this figure is 
            available in the online journal.)}
            \label{loop_evol}
    \end{figure*}
%

In this paper, we analyze a particular dataset of a quiet Sun region close to disk 
center obtained on 25 October 2005 between 13:17 and 13:56\,UT. We focus on the 
time series of high resolution magnetograms from the Michelson Doppler Imager (MDI) 
on board \textit{SoHO}
\citep{1995SoPh..162..129S}                       
and \textit{TRACE} 
\citep{1999SoPh..187..229H}                       
filtergrams taken in the 1216\,\AA~channel. Context images showing the common field of 
view (FoV) in all observations are displayed in 
Fig.\,\ref{context_image}.                        

The MDI longitudinal magnetograms have a FoV of  
614$^{\prime\prime}$\,$\times$\,384$^{\prime\prime}$ (1024\,$\times$\,640\,pixels).
Their spatial resolution in the high-resolution observing mode was 1\farcs25
with a corresponding pixel size of 0\farcs6\,$\times$\,0\farcs6.
The instrumental detection threshold for the magnetic flux density is 
approximately 17\,Mx\,cm$^{-2}$
\citep{1997SoPh..175..329S}.                      
Therefore, all pixels with a signal below this limit were excluded from further 
analysis. The magnetograms were taken with a regular 1 minute cadence, but their 
sequence was interrupted between 13:45\,-\,13:53\,UT. All MDI data were downloaded 
already in the pre-processed form. 

The TRACE data cover a FoV of
384$^{\prime\prime}$\,$\times$\,384$^{\prime\prime}$ (768\,$\times$\,768\,pixels)
with a spatial resolution of 1\farcs0 and a corresponding pixel size of  
0\farcs5.~The sequence of 1216\,\AA~filtergrams 
was taken with a cadence of $\sim$\,45\,s which was interrupted several times
by a cycle of exposures acquired in white-light, 1550\,\AA, 1600\,\AA, 171\,\AA, 
and 196\,\AA~channels. These additional data were taken for context and coalignment 
purposes. All raw images acquired by TRACE were corrected for instrumental 
effects (i.e.~subtracting dark current and pedestal, correcting for exposure 
time, for radiation spikes and for saturated pixels) using the standard procedures 
included in the Solar Software (SSW) IDL routines provided by the TRACE team
\citep{1998SoPh..182..497F}.                      
The data were corrected also for solar rotation. 

Note that the data acquired in the 1216\,\AA~TRACE channel do not include only 
the desired Ly\,$\alpha$ emission, but also signals from UV emissions near 
1550\,\AA~and longer wavelengths.~The reason is a double peak in spectral 
response of the detector with one peak located at $\sim$1216\,\AA~and a second 
peak situated at $\sim$1550\,\AA. The bandwidth of both transmission windows 
is similar and amounts to 84\,\AA~for the one centered at 1216\,\AA.
\citet{1999SoPh..190..351H}                        
described a method how to extract the "pure" Ly\,$\alpha$ emission from observations 
taken in the 1216\,\AA~channel using filtergrams acquired in the 1600\,\AA~channel. 
Because of the absence of a co-temporal sequence of 1600\,\AA~images in our dataset,
we could not apply this method for the present data. The same authors claim that 
only $\sim$60$\%$ of the signal recorded in the 1216\,\AA~channel originates from 
Ly\,$\alpha$ emission. Moreover, the emission recorded in this channel 
covers the temperature range 1.0\,-\,3.0\,$\times$\,10$^{4}$\,K
\citep{1999SoPh..187..229H},                      
i.e., not only the formation temperature of the Ly\,$\alpha$ line which is 
$\sim$2.0\,$\times$\,10$^{4}$\,K 
\citep[e.g.][]{2001ApJ...563..374V}.              
Therefore, we consider the TRACE 1216\,\AA~data only as an approximation 
of chromospheric emission and we do not draw any conclusions about absolute 
intensity changes.                              

A precise coalignment of the MDI and TRACE data had to be performed before the analysis. 
We used white-light images recorded simultaneously with both instruments for this purpose. 
As an independent check, we used the co-spatiality between the brightest chromospheric 
structures visible in the 1216\,\AA~channel (chromospheric network) and magnetic 
features. 
%
   \begin{figure}[!t]
   \centering
   \includegraphics[width=\hsize]{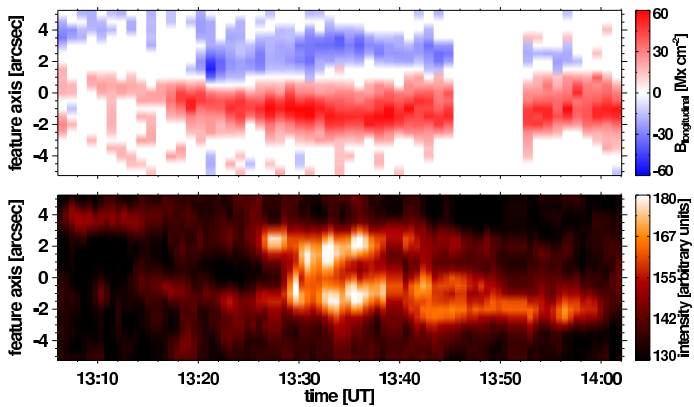}
      \caption{Space-time maps showing the evolution of longitudinal 
               magnetic flux densities and chromospheric emission 
               along the axis connecting the two patches of opposite 
               magnetic polarity of the emerging feature. (A color version 
               of this figure is available in the online journal.)}
         \label{param_evol}
   \end{figure}

\section{Results}

We visually inspected the SoHO/MDI high-resolution magnetograms for emerging magnetic 
dipoles that exhibit also some activity in chromospheric emission. The 
evolution of such a feature emerging at the vicinity of a network boundary is 
presented in Fig.\,\ref{loop_evol}. Note that we used a non-linear time scale for the 
presentation of the sequence of TRACE 1216\,\AA~images with superimposed contours of 
SoHO/MDI magnetograms to provide a better description of the various phases taking 
place during the evolutionary process.

In the first two snapshots (taken at 13:14:15\,UT and 13:16:36\,UT) several 
magnetic elements are visible.~These features represent the pre-existing magnetic 
field forming the network boundary which also remains detectable during and after the 
time span under investigation.~Between 13:17:10\,-\,13:18:20\,UT, a new patch of positive 
polarity and subsequently (at 13:20:05\,UT) also a tiny area of new negative polarity 
become apparent. Finally, at 13:21:15\,UT, the new magnetic feature is well visible. 
Since longitudinal magnetograms are blind to horizontal fields, we were not able to 
correctly describe the complex topology of the emerged feature. However, such 
bipolar events can be best explained either by an emerging $\cap$-like loop or a 
submerging feature with $\cup$-like shape. In both cases, the patches of opposite 
polarity correspond to the footpoints of the emerging feature.

Later on, at 13:25:59\,UT, chromospheric brightenings co-spatial with the new patch 
of negative polarity appear and persist.~After several minutes (at 13:28:32\,UT) 
enhanced chromospheric emission becomes visible also in regions covered by the 
positive-polarity patch.~From 13:30:51\,UT to 13:33:11\,UT enhanced chromospheric 
activity is already apparent along the whole axis of the feature, where we define 
the axis as the straight line connecting the two patches of opposite polarity.
From 13:34:21\,UT until 13:40:42\,UT, the 1216\,\AA~channel emission becomes again 
spatially separated and brightenings related to the negative patch start to weaken 
at the end of this period. Between 13:41:51\,UT and 13:45:21 UT, the northern 
negative-polarity footpoint is still visible in the magnetograms, but does 
not leave longer its detectable fingerprints in chromospheric emission. Finally, at 
13:53:10\,UT and later, there is no clear evidence of the negative footpoint or its 
recurrent appearance (we remind that there is a gap in the magnetograms between 
13:45\,-\,13:53\,UT).  
%
   \begin{figure}[!t]
   \centering
   \includegraphics[width=\hsize]{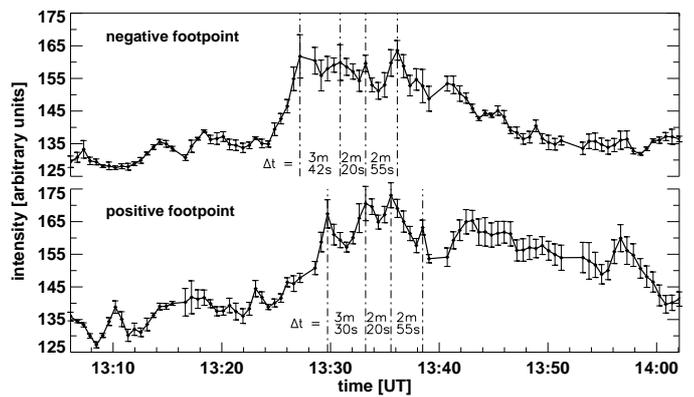}
      \caption{Temporal evolution of the 1216\,\AA~emission at 
               locations of the emerging feature that are associated with 
               strongest photospheric magnetic field ({\em upper curve}: 
               negative polarity; {\em lower curve}: positive polarity). 
               The standard deviation for each intensity value is marked by 
               error-bars. Different values of $\Delta$t depict individual 
               time intervals between consecutive intensity peaks.}
         \label{figure_oscil}
   \end{figure}
%
  
Concerning the southern positive-polarity patch, we note that its interaction and 
consecutive merging with the nearby network boundary complicates its life time estimate 
although significant emission in the co-spatial area in the chromosphere persists even 
after its disappearance. Both interaction and consecutive merging with the network boundary 
can be best seen by means of the contours related to the positive polarity patch in 
Fig.\,\ref{loop_evol} at 13:30:51 UT and later on.

%
\begin{table*}
\caption{Physical parameters of the detected bipolar magnetic feature  
         compared with typical values determined for emerging small-scale 
         magnetic loops by \citet{2009ApJ...700.1391M}.}  
\label{table1}      
\centering                          
\begin{tabular}{l c c c}        
\hline\hline                 
parameter & detected feature & typical loop \\    
          & (positive; negative polarity) &  \\
\hline                        
   longitudinal magnetic flux [Mx] & 1.6$\times 10^{18}$; -8.5$\times 10^{17}$ & 2$\times 10^{16}$\,-\,2$\times 10^{17}$/1.9$\times 10^{18}$\,$^{*}$ \\             
   longitudinal magnetic flux density [Mx\,cm$^{-2}$] & 55; -25 & 20\,-\,40/75.8\,$^{*}$ \\
   distance between the two footpoints & from $\sim$2\farcs5 to $\sim$5\farcs0 & 0\farcs6\,-\,5\farcs5 \\
   lifetime [min] & $\leq$\,36 & 2\,-\,40 \\
   delay between photospheric and chromospheric activity [min] & $\sim$9.0 & 5\,-\,13 \\ 
\hline                       
$^{*}$ - extreme cases                               
\end{tabular}
\end{table*}
%

Figure\,\ref{param_evol} shows the temporal evolution of the longitudinal magnetic flux 
densities at the footpoints {\em (upper panel)} and the chromospheric emission along 
the axis of the emerging magnetic feature {\em (lower panel)}. We calculated the 
longitudinal magnetic flux densities at particular positions across the footpoints by 
averaging over all pixels that are located perpendicular to the loop axis and that do not 
exceed the lateral extension of the two patches of opposite polarity. The changes in flux 
density show a time delay between the two opposite polarity patches and point to an 
asymmetric nature 
of the detected dipole. While the longitudinal magnetic flux densities related to the 
positive polarity patch reach roughly 55\,Mx\,cm$^{-2}$ and remain more or less constant 
during the whole lifetime of the small-scale magnetic feature, the negative polarity 
patch reaches on average only -30\,Mx\,cm$^{-2}$. Nevertheless, the magnetic 
flux densities of both polarities have very similar values at the very beginning of 
the event, when also the negative patch reaches $\sim$\,53\,Mx\,cm$^{-2}$.~Corresponding 
values of the total magnetic flux related to the positive and negative polarity patches
(not shown here) are roughly 1.6$\times 10^{18}$\,Mx and -8.5$\times 10^{17}$\,Mx, 
respectively. As in case of the flux density, also the total flux related to the small-scale 
feature is at least balanced at the beginning of the feature emergence, with an absolute 
value of the total magnetic flux of $\sim$1.5$\times 10^{18}$\,Mx for both 
polarities. Both magnetic flux density and total magnetic flux have been calculated 
within the areas highlighted by yellow contours in Fig.\,\ref{loop_evol}.   

The spatial distance between the footpoints of the emerging feature at photospheric 
level is steadily increasing during first phases of the event from $\sim$2\farcs5 up 
to $\sim$5\farcs0. However, just towards the end of its lifetime the emerging small-scale 
magnetic loop seems to slightly decrease again its axial extension. The lifetime of the 
detected feature can be estimated only very roughly to no more than 36\,minutes because of 
the gap in the MDI data (Figure\,\ref{param_evol}, {\em upper panel}). 

The {\em lower panel} of Fig.\,\ref{param_evol} outlines the delay of $\sim$9\,minutes between 
magnetic feature emergence and chromospheric response, the latter being first visible in the 
region co-spatial to the negative polarity patch. Later on, the chromospheric brightenings 
appear in form of isolated, quasi-periodical bursts, which are visible along the whole axis
connecting the two patches of opposite polarity. The duration of the observed chromospheric 
activity is $\sim$14\,minutes. 

The quasi-periodic nature of the chromospheric emission related to the flux emergence 
is also visible in Fig.\,\ref{figure_oscil}. This plot shows the 1216\,\AA~channel emission 
at locations of the emerging feature associated with strongest magnetic field. 
The increase of the chromospheric signals visible during the main phase of activity 
shows time delays in the range of 2.3 to 3.7 minutes. However we could not assign any 
characteristic period to this oscillatory behaviour. Nevertheless, although the intensity 
enhancements show a time delay between the patches of positive and negative polarity, 
the temporal differences between the particular peaks are the same for both footpoints 
(at least within the temporal resolution of the data), thus increasing the statistical 
significance of these events. 

\section{Discussion}

In Table\,\ref{table1}, we compare the physical parameters of the detected bipolar magnetic 
feature with typical values determined for small-scale magnetic loops emerging from the 
photosphere to higher atmospheric layers 
\citep{2009ApJ...700.1391M}.                      
Note that a proper determination of the magnetic field strength from the derived longitudinal 
magnetic flux densities was impossible because of the missing information about azimuth, 
inclination and filling factor of the magnetic field lines forming the emerging feature. 
Since the magnetic field should be more or less vertical at the footpoints (because of the 
geometry of the loop), the lacking information concerning the inclination is not that crucial. 
If we assume a filling factor close to unity for these areas, then the real magnetic field 
strength is roughly equal to the measured magnetic flux density. However when assuming a much 
smaller filling factor, the real magnetic field strength would increase easily up to kG values. 
Anyway, the comparison of the individual parameters shows that the values estimated for 
the magnetic feature under investigation are consistent with typical characteristics of 
small-scale magnetic loops reaching chromospheric heights. Exceptions are the determined 
longitudinal magnetic flux values in both footpoints which are at least one magnitude larger 
than those of typical loops, thus being rather comparable with extreme cases of the latter 
features. This might be due to the limited spatial resolution of the MDI magnetograms, 
resulting in an overestimation of the total area covered by the footpoints of the emerging 
dipole and the calculated total magnetic flux values. This, together with the fact, that 
the emerging dipole was first detected in the magnetograms and only later also at chromospheric 
heights supports the hypothesis that the observed structure can be identified as a rising 
small-scale magnetic loop, exhibiting significant activity in the upper atmosphere.  

On the other hand, one could object that the almost ideal co-spatiality of the 
patches harboring opposite magnetic polarity with the 1216\,\AA~emissions points 
to a non loop-like shape of the detected feature. The expected quasi-circular 
topology of the detected feature would require that the spatial extent between areas of 
chromospheric emissions should be smaller than the distance between the patches of opposite 
polarity. However,  
\citet{2010ApJ...714L..94M}                        
demonstrated that at the first stages, during the loop emergence into photospheric layers, 
the loop has a flattened geometry and keeps its shape throughout the photosphere. However, 
when passing the transition between photosphere and chromosphere, the magnetic field  
at the footpoints becomes almost vertical and the loop has a $\cap$-like (or arch-like) 
shape. Given this result, the co-spatiality detected in our data even strongens our 
interpretation. 

The axial distance between the footpoints of the loop shows a steady increase in the 
early phase of the event (Fig.\,\ref{param_evol}, {\em upper panel}). After reaching a 
maximum distance, the footpoints remain their position nearly constant to each other until 
the disappearance of the loop. This is a typical behaviour for a small-scale magnetic loop 
within a granule. These loops are first advected by granular flows, and their size increases. 
The almost linear increase of the axial loop dimension indicates that the footpoints 
do not undergo a free random walk. Later, when reaching the intergranular lanes, the footpoints 
remain stable and the loop does not change anymore significantly its axial extension. 
However, a small decrease of the axial loop extension visible after $\sim$13:36\,UT might 
indicate a subsequent submergence of the loop which is also supported by the co-temporal 
weakening of chromospheric emission that vanishes at the end. However, the decrease 
in the axial extension of the loop towards the end of its lifetime could just reflect nothing 
else than the noisy behavior of the SoHO/MDI magnetograms. 

This noisy behaviour also did not allow us to study directly possible oscillations 
of the magnetic flux density at the footpoints.~However, we clearly detected a quasi-periodic 
behavior of the emissions recorded in the 1216\,\AA~channel (Fig.\,\ref{figure_oscil}). We 
estimated the damping time of these fluctuations to $\sim$14\,minutes and their related 
periods to 2.3\,–\,3.7 minutes. These results are in good agreement with findings of 
\citet{2011ApJ...730L..37M}                        
for stronger magnetic patches. The latter authors analyzed high-resolution spectro-polarimetric 
data obtained with SUNRISE/IMaX and found that areas of circular polarization patches, 
containing constant magnetic flux, can oscillate, implying that the magnetic flux density 
fluctuates in antiphase. Typical periods of these oscillations range between 4 and 11 minutes 
in the case of weaker flux patches and between 3 and 5 minutes in case of 
stronger ones. The detected oscillations can be strongly damped or amplified 
within a time interval of 5\,–\,30 minutes. In their paper, the authors guess that these 
oscillations could propagate even up to chromospheric layers in form of waves and dissipate 
their energy through formation of shocks. Thus the pattern detected in the 
1216\,\AA~intensity evolution of our data might also be the result of such events.

Further evidence of oscillations of magnetic flux where found in relation to plages 
and pores 
\citep{2009ApJ...702.1443F}                       
confirming that this phenomenon is also present in other solar features. Also  
\citet{2013A&A...553A..73B}                        
suggest that magnetic heating processes are more important for the chromospheric energy budget 
as commonly assumed.~However, weather acoustic waves or magneto-acoustic waves related with 
e.g. small-scale magnetic loops suffice in heating the solar chromosphere might be answered 
only by future high spatial resolution observations that combine both photospheric 
spectro-polarimetry and chromospheric spectroscopy, e.g. by parallel observations with the 
\textit{GREGOR} Fabry-P\'{e}rot Interferometer 
\citep[GFPI;][]{2012ASPC..463..423P,2012SPIE.8446E..79P,2012AN....333..880P}  
and the planned Blue Imaging Solar Spectrometer 
\citep[BLISS;][]{2012SPIE.8446E..79P,2013arXiv1302.7157P} 
at the 1.5-meter \textit{GREGOR} solar telescope 
\citep{2012AN....333..796S,2012ASPC..463..365S}.  

The small-scale loop under investigation emerged as a dipole which polarities exhibit 
different field strengths most of its lifetime. Thus the total magnetic flux of the 
emerging feature seems to be unbalanced. However, this unexpected result could just reflect 
the difficulties in separating the positive footpoint of the loop from the stronger network 
fields nearby because of its permanent interaction with the latter and its final merge. 
Another possible explanation could be based on the different spatial extension of both 
footpoints. For the positive polarity patch, the magnetic flux could be concentrated within 
a much smaller area, implicating flux densities well above the detection threshold, while 
for the negative polarity the opposite could be the case. Similar conditions can even lead 
to measurements of unipolar fields in quiet Sun    
\citep{2008ApJ...674..520L}.                      
%

\section{Conclusions}

We observed a magnetic feature present in SoHO/MDI magnetograms and TRACE 1216\,\AA~channel 
filtergrams however with a time lack of $\sim$9 minutes in the latter. We identified this 
structure as a small-scale magnetic loop emerging into the photosphere and reaching 
chromospheric heights later on. We interpreted quasi-periodic variations found in the 
chromospheric emission related to the loop as a consequence of magnetic field oscillations 
at the footpoints resulting in a wave propagation towards higher layers. The asymmetric 
appearance of the loop could implicate that the footpoints cover areas of different sizes.    

\begin{acknowledgements}
      This work was supported by the German
      \emph{Deut\-sche For\-schungs\-ge\-mein\-schaft, DFG\/} project
      number BA 1875/7-1, the \emph{Science Grant Agency} - project 
      VEGA 2/0108/12, and the \emph{Slovak Research and Development Agency} 
      under the contract No. APVV-0816-11. 
      \emph{MDI} is part of \emph{SOHO}, the Solar and Heliospheric Observatory, 
      which is a project of international cooperation between ESA and NASA.
      The Transition Region and Coronal Explorer, \emph{TRACE}, is a mission 
      of the Stanford-Lockheed Institute for Space Research, and part of the 
      NASA Small Explorer program.
      We thank Dr. J. Ryb\'{a}k for his help with the data coalignment. 
      The authors thank an anonymous referee for constructive 
      comments and valuable suggestions.
      This research has made use of NASA’s Astrophysics Data System.
\end{acknowledgements}




\end{document}